\documentstyle[12pt,epsf]{article}

\def\beq{\begin{eqnarray}}
\def\eeq{\end{eqnarray}}
\def\ee{\varepsilon}
\def\mprp{\mbox{\tiny $\bot$}}
\def\mprl{\mbox{\tiny $\|$}}

\def\pp{p_{\mbox{\tiny $\|$}}}
\def\zt{X_{\mbox{\tiny $\bot$}}}

\def\Pl{{\cal P}^\lambda}

\def\M{{\cal M}}

\def\lm{\lambda}
\def\half{\frac{1}{2}}
\newcommand{\eq}[1]{(\ref{#1})}
\newcommand{\prp}[1]{#1_{\mbox{\tiny $\bot$}}}
\newcommand{\prl}[1]{#1_{\mbox{\tiny $\|$}}}
\def\HH{H\!\!\left(\frac{4 m^2}{\prl{k}^2}\right)}
\def\HHi{H\!\!\left(\frac{4 m^2}{\prl{(k')^2}}\right)}
\def\HHii{H\!\!\left(\frac{4 m^2}{\prl{(k'')^2}}\right)}
\def\ggg{\gamma \rightarrow \gamma \gamma}
\def\frb{\gamma_{\mprl} \rightarrow \gamma_{\mprl} \gamma_{\mprp}}
\def\alw{\gamma_{\mprl} \rightarrow \gamma_{\mprp} \gamma_{\mprp}}
\def\ff{\Lambda}
\def\tff{\widetilde \Lambda}
\def\1{1 \to 1 \, 2}
\def\2{1 \to 2 \, 2}

\title{
\begin{flushright}
{\normalsize Yaroslavl State University\\
             Preprint YARU-HE-98/03\\
             hep-ph/9804444} \\[5mm]
\end{flushright}
Photon splitting above the pair creation threshold 
in a strong magnetic field}

\author{{M.V.~Chistyakov, A.V.~Kuznetsov and N.V.~Mikheev}\\ [7mm] 
{\small\it Division of Theoretical Physics, Department of Physics,}\\
{\small\it Yaroslavl State University, Sovietskaya 14,}\\
{\small\it 150000 Yaroslavl, Russian Federation}}

\date{}

\begin{document}

\maketitle

\begin{abstract}
The process of photon splitting $\gamma \to \gamma \gamma$ in 
a strong magnetic field is investigated both below and above the pair 
creation threshold. Contrary to the statement by Baier et al., 
the ``allowed'' channel $\alw$ is shown not to be a comprehensive 
description of splitting in the strong field because the ``forbidden'' channel 
$\frb$ is also essential. The partial amplitudes and the splitting 
probabilities are calculated taking account of the photon 
dispersion and large radiative corrections near the resonance. 
\end{abstract}

\vspace{10mm}

\centerline{\it Submitted to Physics Letters B}

Nowadays it is generally recognized that an external 
elec\-tro\-mag\-ne\-tic field plays the role of an active medium. 
It can induce new 
interactions and can also change the dispersion properties of particles, thus 
allowing processes which are forbidden in vacuum. The photon splitting process 
$\gamma \to \gamma \gamma$ could be a prominent example of the magnetic field 
effect. 

The theoretical study of this process has a rather long 
history~\cite{early,Ad71,early2}. Recent progress in astrophysics 
has drawn attention again to the photon splitting induced by a magnetic field. 
This process may explain the peculiarity of the spectra of the 
observed $\gamma$-bursts~\cite{Bar95,HBG}. The origin of these $\gamma$-bursts 
is not clear yet. Some models exist where astrophysical cataclysms like 
a supernova explosion or a coalescence of neutron stars could be the sources 
of such $\gamma$-bursts. It is generally supposed that these objects have 
strong magnetic fields $ \sim 10^{16} - 10^{17}\, G$~\cite{mag} 
much greater than the Schwinger value, 
$B_e = m^2/e \simeq 4.41 \cdot 10^{13} G$ (hereafter $m$ is the electron 
mass, $e > 0$ is the elementary charge). 

In our opinion, continued investigation of the process
$\gamma \to \gamma \gamma$ is important not only because of its 
possible applications, but to also gain an improved understanding of 
radiative processes in intense external fields. 
The study of the photon splitting $\gamma \to \gamma \gamma$ in a strong 
magnetic field~\cite{Baier,Ad96,W} has so far considered only the 
collinear limit of the process, when the only allowed 
transition with respect to photon polarizations is, $\alw$ (in Adler's 
notation~\cite{Ad71}). However, photon dispersion in a strong 
magnetic field, $B \gg B_e$, leads to significant deviations 
from the collinearity of the kinematics of this process. 
This is due to the fact that the eigenvalues of the 
photon polarization operator (the photon effective mass squared) 
become large near the so-called cyclotron 
resonances~\cite{Shab}. The lowest of them is only essential in a strong 
field and corresponds to the pair creation 
threshold $\omega_0 = 2 m$ (in the frame where the photon momentum is 
perpendicular to the field direction). 
A photon of the $\perp$ mode acquires 
in this region a significant effective mass, $k^2 = \omega^2 - {\bf k}^2 < 0, 
\; |k^2| \sim \omega^2$, 
and this defines the kinematics of the photon splitting, 
which is far from collinearity. 

On the other hand, a large value of the polarization operator near the 
resonance requires taking account of large radiative 
corrections which reduce to a renormalization of the photon 
wave-function  
\beq
\ee_{\alpha}^{(\lm)} \to \ee_{\alpha}^{(\lm)} \sqrt{Z_\lm}, \quad 
Z^{-1}_\lm = 1 - \frac{\partial \Pl}{\partial \omega^2}, \quad 
\lm = \parallel, \perp. 
\eeq    

\noindent Here 
$\ee_\alpha^{(\mprl)}$, $\ee_\alpha^{(\mprp)}$ are the polarization 
four-vectors of the photon modes and $\Pl$ are the eigenvalues of the 
photon polarization operator, corresponding to these modes~\cite{Shab}.

Both the effect of noncollinearity and 
radiative corrections have not, so far, been taken into account. 
For example, in the paper by Baier et al.~\cite{Baier} the amplitude 
and the probability of the photon splitting were obtained for 
the photon energies up to the threshold $\omega_0 = 2 m$. Strictly speaking, 
that result is incorrect because the noncollinearity of the photon momenta 
and the photon wave-function renormalization 
were not considered. In fact, their amplitude describes the photon 
splitting correctly only in the limit $\omega \ll m$. The same criticism is 
appropriate for the paper~\cite{W}. 

It should be stressed also, that the channel $\alw$ is not adequate to 
describe photon splitting because of the noncollinearity of the photon momenta.
In particular, the transition $\frb$ forbidden in the 
collinear limit, gives an essential contribution into the splitting 
probability, contrary to the statement by Baier et al.~\cite{Baier}. 

In this paper, photon splitting in a strong magnetic field is 
investigated both below and above the pair-creation threshold, 
with an emphasis on the noncollinearity of the kinematics, 
and taking account of large radiative corrections. 
The splitting probability of a photon with high energy is of particular 
importance for calculations of the spectra of strongly magnetized 
cosmic objects. 

The process of photon splitting in an external field is depicted by two 
Feynman diagrams, see Fig.1. The electron propagator in a magnetic field 
could be presented in the form 
\beq
S(x,y) &=& e^{\mbox{\normalsize $i \Phi (x,y)$}}\, \hat S(x-y),\label{Sxy}\\
\Phi(x,y) &=& - e \int \limits^y_x d\xi_\mu \left [ A_\mu(\xi) + \half 
F_{\mu \nu}(\xi - y)_\mu \right ],\label{Phi}
\eeq

\noindent where $A_\mu$ is a 4-potential of the uniform magnetic field. 
The translational invariant part $\hat S(x-y)$ of the propagator has several 
representations. It is convenient for our purposes to take it in the form of 
a partial Fourier integral expansion
\beq 
\hat S(X) &=& - \frac{i}{4 \pi} \int \limits^{\infty}_{0} 
\frac{d\tau}{th \tau}\,
\int \frac{d^2 p}{(2 \pi)^2}
\Biggl \{ 
[\prl{(p \gamma)} + m]\Pi_{-}(1 + th \tau) +
\nonumber\\
&+& [\prl{(p \gamma)} + m]\Pi_{+}(1 - th \tau) 
- \prp{(X \gamma)}\; \frac{i eB}{2\, th \tau} (1 - th^2 \tau) 
\Biggr \} 
\times \label{Sx}\\
&\times& \exp\left(- \frac{eB \zt^2}{4\, th \tau} - 
\frac{\tau(m^2 - \pp^2)}{eB} - i \prl{(pX)} \right), 
\nonumber\\[3mm]
d^2 p &=& dp_0 dp_3, \quad \Pi_\pm = \half (1 \pm i \gamma_1 \gamma_2),
\quad \Pi^2_\pm = \Pi_\pm, \quad [\Pi_\pm, \prl{(a \gamma)}] = 0,
\nonumber
\eeq
 
\noindent where $\gamma_\alpha$ are the Dirac matrices in the standard 
representation, the four-vectors with the indices $\bot$ and $\parallel$ 
belong 
to the (1, 2) plane and the Minkowski (0, 3) plane correspondingly, when the 
field $\bf B$ is directed along the third axis. Then for arbitrary 4-vectors 
$a_\mu$, $b_\mu$ one has
\beq
\prp{a} &=& (0, a_1, a_2, 0), \quad  \prl{a} = (a_0, 0, 0, a_3), \nonumber \\
\prp{(ab)} &=& (a \ff b) =  a_1 b_1 + a_2 b_2 , \quad 
\prl{(ab)} = (a \tff b) = a_0 b_0 - a_3 b_3, \nonumber 
\eeq

\noindent where the matrices are introduced
$\ff_{\alpha \beta} = (\varphi \varphi)_{\alpha \beta}$,\,  
$\tff_{\alpha \beta} = 
(\tilde \varphi \tilde \varphi)_{\alpha \beta}$, connected by the relation 
$\tff_{\alpha \beta} - \ff_{\alpha \beta} = 
g_{\alpha \beta} = diag(1, -1, -1, -1)$, 
$\varphi_{\alpha \beta} =  F_{\alpha \beta} /B,\; 
{\tilde \varphi}_{\alpha \beta} = \frac{1}{2} \varepsilon_{\alpha \beta
\mu \nu} \varphi_{\mu \nu}$ are the dimensionless tensor of the external 
magnetic field and the dual tensor, 
$(a \ff b) = a_\alpha \ff_{\alpha \beta} b_\beta$.

In spite of the translational and gauge noninvariance of the phase 
$\Phi(x, y)$ in the propagator~\eq{Sxy}, the total phase of three propagators 
in the loop of Fig.1 is translational and gauge invariant
\beq 
\Phi(x, y) + \Phi(y, z) + \Phi(z, x) = - \frac{e}{2} (z - x)_\mu 
F_{\mu \nu} (x - y)_\nu. \nonumber
\eeq

The amplitude of the photon splitting $\ggg$ takes the form
\beq
\M &=& e^3 \int d^4 X\, d^4 Y\, Sp \{\hat \ee(k) \hat S(Y) \hat \ee(k'') 
\hat S(-X-Y) \hat  \ee(k') \hat S(X)\} \times \nonumber \\
&\times& e^{- i e\,(X F Y)/2}\; e^{i (k' X - k'' Y)} + 
(\ee(k'), k' \leftrightarrow \ee(k''), k''),
\label{M}
\eeq

\noindent where $k_\alpha = (\omega, {\bf k})$ is the 4-vector of the momentum 
of initial photon with the polarization vector $\ee_\alpha$, $k'$ and $k''$ 
are the 4-momenta of final photons, $X = z - x, \, Y = x - y$. 

Substitution of the propagator~\eq{Sx} into the amplitude~\eq{M} leads to 
a very cumbersome expression in a general case. Relatively simple results 
were obtained only in the two limits of a weak field~\cite{Ad71} and of the 
strong field with collinear kinematics~\cite{Baier}. 

It is advantageous to use the asymptotic expression of the electron 
propagator for an analysis of the amplitude~\eq{M} in the strong field without 
the collinear approximation. This asymptotic could be easily derived from 
Eq.~\eq{Sx} by evaluation of the integral over $\tau$ in the limit
$eB /\vert m^2 - \prl{p}^2 \vert \gg 1$. In this case the propagator 
takes the simple form
\beq
\hat S(X) \simeq S_a(X) = \frac{i eB}{2 \pi} \exp (- \frac{eB \zt^2}{4}) 
\int \frac{d^2 p}{(2 \pi)^2}\, \frac{\prl{(p\gamma)} + m}{\prl{p}^2 - m^2}
\Pi_{-}e^{-i \prl{(pX)}}, 
\label{Sa}
\eeq

\noindent which was obtained for the first time in Ref.~\cite{Skob}. 
Substituting this in Eq.~\eq{M} and integrating, one would expect to obtain 
the amplitude which depends linearly on the field strength, namely, as 
$B^3/B^2$, where $B^2$ in the denominator arises from the integration over 
$d^2 X_{\mprp} d^2 Y_{\mprp}$. However, two parts 
of the amplitude~\eq{M} cancel each other exactly. Thus, the asymptotic form 
of the electron propagator~\eq{Sa} only shows that the 
linear-on-field part of the amplitude is zero and provides no way of 
extracting the next term of expansion over the field strength. 

As the analysis shows, that could be done by the insertion of two 
asymptotic~\eq{Sa} and one exact propagator~\eq{Sx} into the amplitude~\eq{M}, 
with all interchanges. It is worthwhile now to turn from the general 
amplitude~\eq{M} to the partial amplitudes corresponding to definite 
photon modes, $\parallel$ and $\perp$, which are just the stationary photon 
states with definite dispersion relations in a 
magnetic field. There are 6 independent amplitudes and only two of them are 
of physical interest. We have obtained the following expressions, to the terms 
of order $1/B$ 
\beq
\M_{\mprl \to \mprl \mprp} &=& 
i 4 \pi \left(\frac{\alpha}{\pi} \right)^{3/2} \, 
\frac{(k' \varphi k'')(k' \tilde \varphi k'')}
{[
\prl{(k')^2}\prp{(k'')^2}\prp{k^2}
]^{1/2}}\; \HHi, 
\label{Mfrb} \\
\M_{\mprl \to \mprp \mprp} &=& 
i 4 \pi \left(\frac{\alpha}{\pi} \right)^{3/2} \, 
\frac{(k' \tff k'')}
{[\prl{(k')^2}\prl{(k'')^2}\prp{k^2}]^{1/2}}
\Biggl \{ (k \ff k'') \HHi 
\nonumber \\
&+& (k \ff k') \HHii \Biggr \}, 
\label{Malw} 
\eeq

\noindent where 
\beq
&&H(z)=\frac{z}{\sqrt{z - 1}} \arctan \frac{1}{\sqrt{z - 1}} - 1, 
\ z > 1,
\nonumber \\
\label{Hz}\\ 
&&H(z) = - \half \left ( \frac{z}{\sqrt{1-z}}
\ln \frac{1 + \sqrt{1-z}}{1 - \sqrt{1-z}} + 2 + 
i \pi \frac{z}{\sqrt{1-z}} \right ), \ z < 1. 
\nonumber
\eeq
   
\noindent As for remaining amplitudes, we note that $\M_{\mprl \to \mprl 
\mprl}$ 
is equal to zero in this approximation. On the other hand, the photon of the 
$\perp$ mode due to its dispersion can split into two photons only in the 
kinematic region $\prl{k^2} > 4 m^2$ where the tree-channel 
$\gamma_{\mprp} \to e^+ \,e^-$~\cite{Klep} strongly dominates.

Thus we will analyse further the photon splitting of the $\|$ mode 
in the region $\prl{k^2} < (m + \sqrt{m^2 + 2 e B})^2$ where the tree-channel 
$\gamma_{\mprl} \to e^+ \,e^-$ does not exist. 
In the formal limit of the collinearity of the photon momenta, the amplitude 
$\M_{\mprl \to \mprl \mprp}$ goes to zero while the amplitude 
$\M_{\mprl \to \mprp \mprp}$ coincides with the amplitude obtained in 
Ref.~\cite{Baier}. 

Although the process involves three particles, its amplitude is not a 
constant, 
because it contains the external field tensor in addition to the photon 
4-momenta. The general expression for the splitting probability can be 
written in the form
\beq
&&W_{\lm \to \lm' \lm''} = 
\frac{g}{32 \pi^2 \omega} \int \vert \M_{\lm \to \lm' \lm''} \vert^2
Z_\lm Z_{\lm'} Z_{\lm''} \times \nonumber \\
&& \times \ \delta(\omega_\lm({\bf k}) - \omega_{\lm'}({\bf k'}) - 
\omega_{\lm''}({\bf k} - {\bf k'})) 
\frac{d^3 k'}{\omega_{\lm'} \omega_{\lm''}},
\label{defW}
\eeq

\noindent where the factor $g = 1 - {1 \over 2}\delta_{\lm' \lm''}$ is 
inserted to account for possible identity of the final photons. The factors 
$Z_\lm$ account for the large radiative corrections which reduce to the 
wave-function renormalization of a real photon with definite dispersion 
$\omega = \omega_\lm({\bf k})$. 
The integration over phase space of two final photons in Eq.~\eq{defW} 
has to be performed using the photon energy dependence on the 
momenta, $\omega = \omega_\lm({\bf k})$, which can be found from 
the dispersion equations
$$\omega_\lm^2({\bf k}) - {\bf k}^2 - {\cal P}^\lm = 0, $$ 
where ${\cal P}^\lm$ are the eigenvalues of the 
photon polarization operator~\cite{Shab}. 
In the limit of a strong field and in the kinematic region 
$\prl{k}^2 = \omega^2 - k_3^2 \ll e B$ one obtains
\beq
{\cal P}^{\mprl} & \simeq & - \frac{\alpha}{3 \pi} k^2_{\mprl},\\
{\cal P}^{\mprp} & \simeq & - \frac{2 \alpha}{\pi} \;e B \;\HH, 
\eeq

\noindent where $H(z)$ is the function defined in Eq.~\eq{Hz}. 

A calculation of the splitting probability~\eq{defW} is rather complicated 
in the general case.  
In the limit $m^2 \ll \omega^2 \sin^2 \theta \ll eB$,
where $\theta$ is an angle between the initial photon momentum
$\bf k$ and the magnetic field direction, we derive the 
following analytical expression for the probability of the channel $\frb$:
\beq
W_{\mprl \to \mprl \mprp} 
&\simeq& \frac{\alpha^3 \omega \sin^2 \theta}{16} (1 - x)
[1 - x + 2 x^2 + 2(1 - x)(1 + x)^2 \ln (1 + x) -
\nonumber\\[2mm]
&-& 2 x^2 \,\frac{2-x^2}{1-x} \, \ln \frac{1}{x}], \qquad
x = \frac{2 m}{\omega \sin \theta} \ll 1. 
\label{Wa}
\eeq

Within the same approximation we obtain the spectrum of 
final photons in the frame where the initial photon momentum is orthogonal 
to the field direction: 
\beq
&&\frac{d W_{\mprl \to \mprl \mprp}}{d \omega'} \simeq \frac{\alpha^3}{2}\cdot
\frac{\sqrt{(\omega - \omega')^2 - 4 m^2}}
{\omega' + \sqrt{(\omega' - \omega)^2 - 4 m^2}}, \label{spektr}\\
&&\frac{\omega}{2} - \frac{2 m^2}{\omega} < \omega' < \omega - 2 m, \nonumber
\eeq

\noindent where $\omega, \omega'$ are the energies of the initial and final 
photons of the $\parallel$ mode. 

We have made numerical calculations of the process probabilities 
for both channels, which are valid in the limit
$\omega^2 \sin^2 \theta \ll eB$. 
Our results are represented in Figs.~2,3. 
The photon splitting probabilities below and near the pair-creation threshold 
are depicted in Fig.~2. In this region the channel $\alw$ (allowed in the 
collinear limit) is seen to dominate the channel $\frb$ (forbidden in 
this limit). For comparison we show here the probability 
obtained without considering the noncollinearity of the kinematics and 
radiative corrections 
(the dotted line $\it 3$) which is seen to be inadequate. 
For example, this probability becomes infinite just above the threshold. 
As is seen from Fig.~3, both channels give essential contributions 
to the probability at high photon energies, with the 
``forbidden'' channel dominating. 
It should be stressed that taking account of the photon polarization 
leads to the essential dependence of the splitting probabilities on the 
magnetic field, while the amplitudes~\eq{Mfrb}, \eq{Malw} do not depend 
on the field strength value. 

In conclusion, we have investigated the photon splitting $\ggg$ below and 
above the pair-creation threshold in the strong magnetic field, 
$B \gg B_e$. We have found the amplitudes and the probabilities of two 
channels, $\alw$ and $\frb$, taking account of the essential influence 
of the field on the process kinematics, and of large radiative corrections. 
The photon spectrum and the splitting probability are calculated 
analytically for the dominating channel $\frb$ at high energies of the 
initial photon. 
The splitting probabilities are calculated numerically for both channels.
The results should be used for quantitative studies of the role of the 
exotic field-induced process of photon splitting in high-energy 
astrophysics.

\bigskip

\noindent 
{\bf Acknowledgements}  

The authors are grateful to V.A.~Rubakov for useful discussions of a 
problem of large radiative corrections in the vicinity of a cyclotron 
resonance, and to M.Yu.~Borovkov for help in preparing the code for 
numerical analysis.
This work was supported in part by the INTAS Grant N~96-0659  
and by the Russian Foundation for Basic Research Grant N~98-02-16694.

\newpage

\thispagestyle{empty}

\centerline{\bf Figure captions}

\centerline{to the paper: M.V.~Chistyakov et al., ``Photon splitting \dots''}

\vspace{5mm}

\begin{itemize}

\item[\bf Fig. 1] 
The Feynman diagram for photon splitting in a magnetic field.
The double line corresponds 
to the exact propagator of an electron in an external field.

\item[\bf Fig. 2] 
The dependence of the probability of photon splitting $\ggg$ on 
energy, below and near the pair-creation threshold: 
$\it 1a, 1b$ -- for the ``forbidden'' channel $\frb$ and for 
the magnetic field strength $B = 10^2\,B_e$ and $10^3\,B_e$, correspondingly;
$\it 2a, 2b$ -- for the ``allowed'' channel $\alw$
for the field strength $B = 10^2\,B_e$ and $10^3\,B_e$; 
$\it 3$ -- for the channel $\alw$ in the collinear limit without taking 
account of large radiative corrections. 
Here $W_0 = (\alpha/\pi)^3 \,m$.

\item[\bf Fig. 3] 
The probability of photon splitting above the pair-creation threshold: 
$\it 1a, 1b$ -- for the ``forbidden'' channel $\frb$ and for 
the magnetic field strength $B = 10^2\,B_e$ and $10^3\,B_e$, correspondingly;
$\it 2a, 2b$ -- for the ``allowed'' channel $\alw$
for the field strength $B = 10^2\,B_e$ and $10^3\,B_e$.

\end{itemize}

\newpage
\thispagestyle{empty}

\begin{figure}[ht]
\epsffile[140 480 0 800]{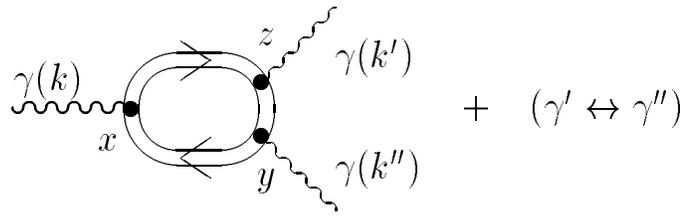}
\caption{M.V.~Chistyakov et al., ``Photon splitting \dots''}
\end{figure}

\newpage
\thispagestyle{empty}

\begin{figure}[ht]
\epsffile[140 320 0 700]{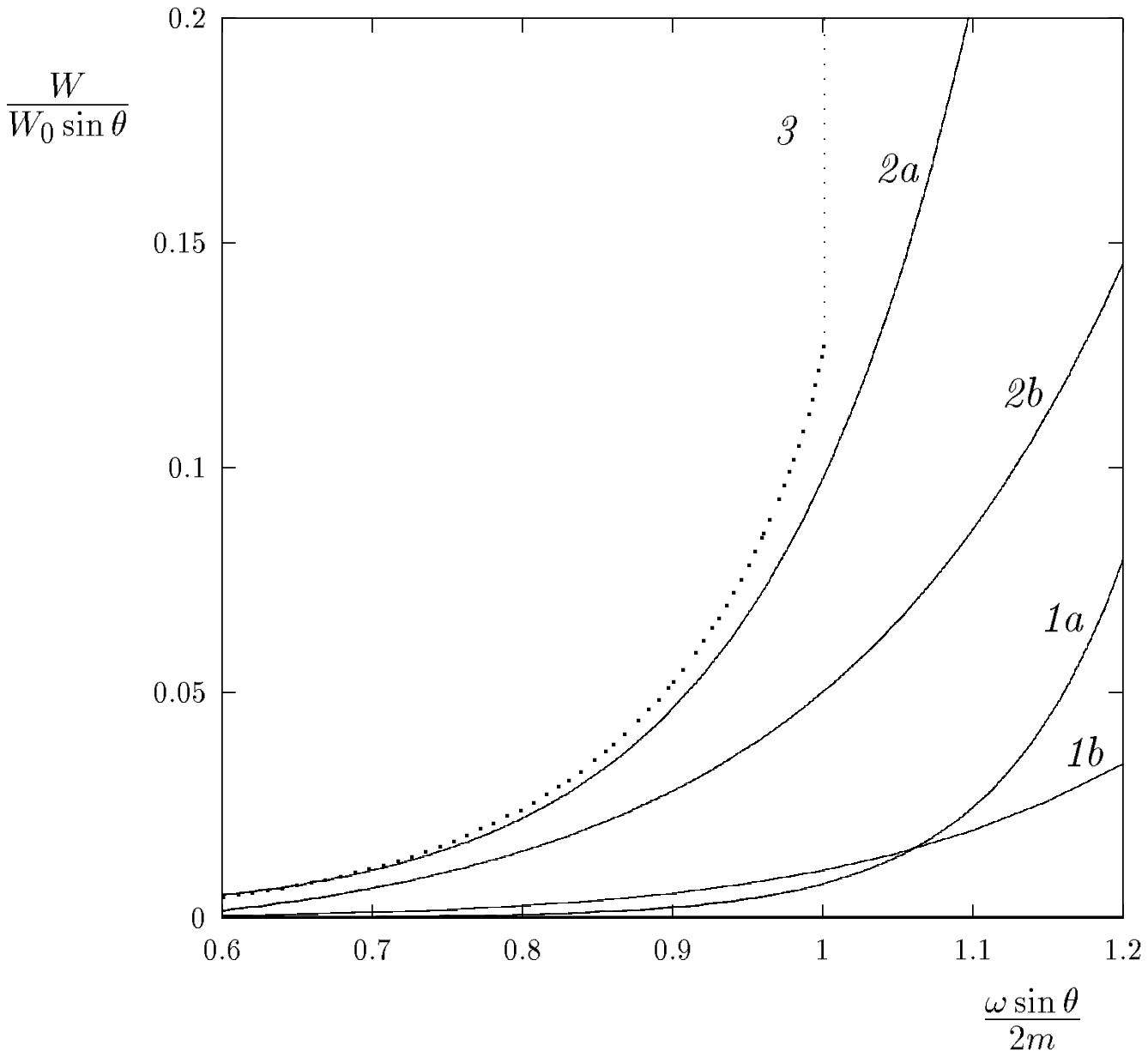}
\caption{M.V.~Chistyakov et al., ``Photon splitting \dots''}
\end{figure}

\newpage
\thispagestyle{empty}

\begin{figure}[ht]
\epsffile[140 320 0 700]{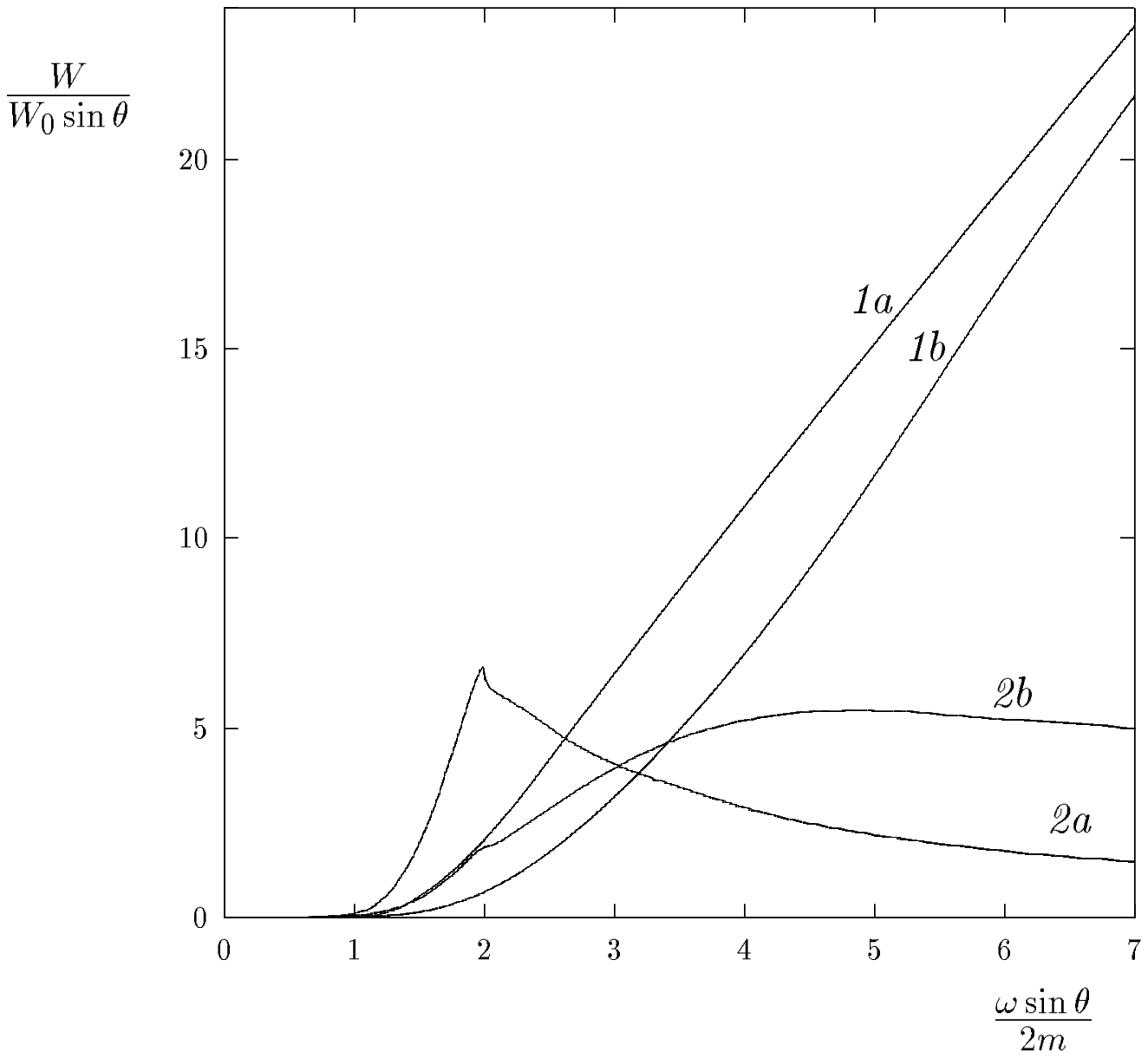}
\caption{M.V.~Chistyakov et al., ``Photon splitting \dots''}
\end{figure}

\end{document}